\documentclass[twocolumn,showpacs,amsmath,amssymb]{revtex4-1}

\usepackage{hyperref}
\usepackage{slashed}

\newcommand{\fant}[1]{\phantom{#1}}
\newcommand{\be}{\begin{equation}}
\newcommand{\ee}{\end{equation}}
\newcommand{\wdg}{\wedge}
\newcommand{\ot}{\otimes}

\begin{document}
\title{Impulsive gravitational waves in general massive 3D gravity}
\author{Ahmet Baykal}
\affiliation{ Department of Physics, Faculty of Arts and Sciences, Ni\u gde \"Omer Halisdemir University,  
Merkez Yerle\c{s}ke, 51240 Ni\u gde, TURKEY}
\author{Tekin Dereli}
\affiliation{Department of Physics, College of Sciences, Ko\c{c} University,  34450 Sar\i yer, \.Istanbul, TURKEY}
\date{\today}

\begin{abstract}
Impulsive, nondiverging,  Petrov-Segre type-N gravitational wave solutions to a general massive three-dimensional gravity 
in the de Sitter, anti-de Sitter and flat Minkowski backgrounds are constructed in a unified manner by using the exterior  algebra of differential forms.
\end{abstract}

\maketitle

\section{Introduction}\label{intro}

General relativity has no propagating degree of freedom in three dimensions. This shortcoming is remedied by complementing the Einstein-Hilbert action
by the Chern-Simons term leading to a parity-violating, topologically  massive gravity (TMG) \cite{DJT, DJT2,DJT3}. More recently, a new massive gravity (NMG) model involving a particular combination of quadratic curvature terms  leading to a second order trace has been introduced by Bergshoeff, Hohm and Townsend \cite{BHT1}.
The NMG model is parity preserving and has only a massive spin-2 particle propagating in flat background with $\mp2$ helicity modes of equal masses.
Furthermore, NMG is unitary at the tree level \cite{gullu-tekin}.

In this paper, a new family of gravitational wave solutions to the general massive gravity \cite{BHT2} extended by a cosmological parameter (CGMG)  is presented.
The CGMG model is governed by the Lagrangian 3-form density
\begin{eqnarray}
&&L_{CGMG}
=
\left[
\sigma R-2\lambda m^2+\frac{1}{m^2}\left(R_{ab}R^{ab}-\frac{3}{8}R^2\right)
\right]*1
\nonumber\\
&&\hspace{1.3cm}+
\frac{1}{\mu}\left(\omega^{a}_{\fant{a}b}\wdg d \omega^{b}_{\fant{a}a}
+
\frac{2}{3}\omega^{a}_{\fant{a}b}\wdg \omega^{b}_{\fant{a}c}\wdg \omega^{c}_{\fant{a}a}\right)\label{CGMG-lag-3form}
\end{eqnarray}
where $m$ and $\mu$ are the mass parameters that appear as coupling constants in the quadratic curvature  and Lorentz Chern-Simons parts of the Lagrangian. $m^2\lambda$ is a cosmological parameter, and the constant $\sigma$ is introduced to register the sign of the mass term of a free spin-2 field. The remaining notation is fully introduced in the section below.

The divergence-free gravitational wave solutions to TMG  that belong to the Kundt family have been thoroughly discussed in 
\cite{chow-pope-sezgin1,chow-pope-sezgin2} classifying all known solutions in just three families.

The gravitational wave solutions to three-dimensional gravitational models have been studied recently  from various perspectives,
\cite{macias-camacho}. Ahmedov and Aliev \cite{AA3,AA1,AA2} introduced a particularly convenient technique to generate the Petrov-Segre type-N solutions to NMG from
the TMG solutions. Ay\`on-Beato et al studied the Ads-wave solutions to NMG \cite{ayon-beato-giribet-hassaine}.
 More recently,  G\"urses et al. \cite{gurses-sisman-tekin1,gurses-sisman-tekin2, gurses-sisman-tekin3, gurses-sisman-tekin4}
also studied solutions to general three-dimensional gravitational models in an exhaustive manner.

The paper is organized as follows.
The required geometrical techniques are developed from  scratch  using the algebra of differential forms \cite{dereli-tucker,obukhov} in the following section. The CGMG field equations that follow from (\ref{CGMG-lag-3form}) are presented in Sec. III. The linearized form of the field equations is also briefly discussed at the end of Sec. III. The metric ansatz along with some of its geometric properties have been introduced in Sec. IV. In the subsequent sections, the differential equation for the profile function is derived after giving the curvature expressions of the ansatz.

\section{Geometrical preliminary}

The geometrical setting for the seminull coframe basis with varying conventions has been introduced  previously
\cite{hall-morgan-perjes,nutku-aliev,hall-capocci,milson-wylleman} in the literature. The notation and most of the conventions used below follow
those  of Aliev  and Nutku introduced in a  spinor formulation of TMG.

In what follows, a seminull coframe
$\{\theta^a\}=\{k, l, m\}$  for $a=0,1,2$ is made use of throughout. The Hodge dual of unity defines the volume form, $*1=k\wdg l\wdg m$, as the exterior product of basis 1-forms.
In terms of the seminull coframe basis, the metric takes the form
\be\label{metric-general-seminull-form}
g=-k\ot l-l\ot k+m\ot m.
\ee
The indices are raised/lowered by $\eta_{ab}$ of the seminull coframe.
The definition of the invariant volume 3-form and the metric coefficients $-\eta_{01}=-\eta_{10}=\eta_{22}=1$ relative to the seminull coframe, are sufficient to
define the Hodge dual of an arbitrary form. To define the Hodge dual on an arbitrary form it suffices to have the Hodge duals 
\be
*k=k\wdg m,\quad *l=-l\wdg m,\quad *m=k\wdg l.
\ee
$i_X$ is the operator denoting a contraction of a $p$-form with a vector field $X$. The set of frame fields $\{e_a\}$ for $a=0, 1, 2$, which are metric dual to the basis coframe 1-forms, are
usually denoted as
\be
e_0=-\tilde{l}\equiv-\Delta, \quad e_1=-\tilde{k}\equiv -D, \quad e_2=\tilde{m}\equiv \delta.
\ee
The isomorphism indicated by a $\tilde{\phantom{a}}$ over  a 1-form field  giving the associated vector field  amounts to
the raising of  a covariant index by means of the inverse metric.
In what follows, a contraction with respect to a basis frame field $e_a$ is abbreviated as $i_a$.

In terms of the frame fields, the exterior derivative can be written as
\be
d=-k \Delta-lD+m\delta
\ee
acting on the scalars. 

Relative to a seminull coframe, there are three independent Levi-Civita connection 1-forms $\omega^{a}_{\fant{a}b}$, namely,
$\omega^{0}_{\fant{a}0}$, $\omega^{0}_{\fant{a}2}$ and $\omega^{1}_{\fant{a}2}$ and as a consequence of the metricity  property, one has
$\omega^{0}_{\fant{a}0}=-\omega^{1}_{\fant{a}1}$, $\omega^{0}_{\fant{a}2}=\omega^{2}_{\fant{a}1}$ and $\omega^{2}_{\fant{a}1}=\omega^{0}_{\fant{a}2}$.
In terms of the three independent Levi-Civita connection forms, the Cartan's first structure equations 
\be
d\theta^a+\omega^{a}_{\fant{a}b}\wdg\theta^b=0
\ee
for $a=0, 1, 2$ read explicitly
\begin{eqnarray}
&dk+\omega^{0}_{\fant{a}0}\wdg k+\omega^{0}_{\fant{a}2}\wdg m=0,
\label{se1-1}\\
&dl-\omega^{0}_{\fant{a}0}\wdg k+\omega^{1}_{\fant{a}2}\wdg m=0,
\label{se1-2}\\
&dm+\omega^{1}_{\fant{a}2}\wdg k+\omega^{0}_{\fant{a}2}\wdg l=0,\label{se1-3}
\end{eqnarray}
where the numerical tensorial indices exclusively refer to the seminull coframe.
In terms of the corresponding curvature 2-forms, the Cartan's second structure equations
\be
\Omega^{a}_{\fant{a}b}
=
d\omega^{a}_{\fant{a}b}+\omega^{a}_{\fant{a}c}\wdg \omega^{c}_{\fant{a}b},
\ee
for $a=0, 1,2$ read
\begin{eqnarray}
&\Omega^{0}_{\fant{a}0}
=
d\omega^{0}_{\fant{a}0}+\omega^{0}_{\fant{a}2}\wdg \omega^{1}_{\fant{a}2},
\label{se2-1}\\
&\Omega^{0}_{\fant{a}2}
=
d\omega^{0}_{\fant{a}2}+\omega^{0}_{\fant{a}0}\wdg \omega^{0}_{\fant{a}2},
\label{se2-2}\\
&\Omega^{1}_{\fant{a}2}
=
d\omega^{1}_{\fant{a}2}-\omega^{0}_{\fant{a}0}\wdg \omega^{1}_{\fant{a}2}.\label{se2-3}
\end{eqnarray}

The Ricci 1-forms $R_a=R_{ab}\theta^b$ can be defined in terms of the contraction $R_a=i_b\Omega^{b}_{\fant{a}a}$.
Accordingly, the scalar curvature can be expressed as $R=i_aR^a$. In three dimensions, the Einstein 1-forms $G_{a}\equiv G_{ab}\theta^b$ can be defined by the relation $*G^a=-\frac{1}{2}\Omega_{bc}*(\theta^{a}\wdg \theta^{b}\wdg \theta^{c})$, which also leads to $G^{a}_{\fant{b}b}=\frac{1}{4}\delta^{acd}_{bmn}R^{mn}_{\fant{mn}cd}$ in terms of the Riemann tensor $R^{mn}_{\fant{mn}cd}$ with  $\Omega^{a}_{\fant{a}b}=\frac{1}{2}R^{a}_{\fant{a}bcd}\theta^{b}\wdg \theta^d$ and the generalized Kronecker delta. In the present formalism, one has the convenient relations
\be\label{3d-einstein-forms}
*G^0=-\Omega^{0}_{\fant{a}2},\quad
*G^1=\Omega^{1}_{\fant{a}2},\quad
*G^2=\Omega^{0}_{\fant{a}0},
\ee
between the Einstein and the curvature forms.

The Cotton 2-forms $C^a=\frac{1}{2}C^{a}_{\fant{a}bc}\theta^b\wdg \theta^c$ can be defined as the covariant exterior derivative of the Schouten 1-forms
$Y^a=R^a-\frac{1}{4}R\theta^a$ as $C^a\equiv DY^a=dY^a+\omega^{a}_{\fant{a}b}\wdg Y^b$
and, relative to the seminull coframe these equations explicitly read
\begin{eqnarray}
&C^0
=
dY^0+\omega^{0}_{\fant{a}0}\wdg Y^0+\omega^{0}_{\fant{a}2}\wdg Y^2,
\label{cotton-sn0}\\
&C^1
=
dY^1-\omega^{0}_{\fant{a}0}\wdg Y^1+\omega^{1}_{\fant{a}2}\wdg Y^2,
\label{cotton-sn1}\\
&C^2
=
dY^2+\omega^{1}_{\fant{a}2}\wdg Y^0+\omega^{0}_{\fant{a}2}\wdg Y^1.\label{cotton-sn2}
\end{eqnarray}
Because the Cotton 2-forms $C^a$ can be derived from the Chern-Simons Lagrangian density 3-form by a coframe variational derivative, they are covariantly constant 2-forms. In three dimensions the Weyl 2-form is not defined and the conformal flatness requires the vanishing of Cotton 2-forms  in three dimensions.

The geometrical formulas developed above are in sufficient generality and allow one to cast any three-dimensional theory into a seminull coframe in terms of differential forms. For example,  Eqs. (\ref{3d-einstein-forms}) together with Eqs. (\ref{cotton-sn0}-\ref{cotton-sn2})  can be used to write out the TMG  equations using the spinorial components given in scalar form in, e.g. Ref. \cite{nutku-aliev}, along with the appropriate changes in the conventions above.

Finally, note that in the present notation, the Cotton-York tensor $C^{ab}$  relative to an orthonormal/seminull coframe can be related to the Cotton 2-forms defined above by the formula
\be\label{cotton-york}
C^{ab}
\equiv
i^a*C^b
=
*(DY^b\wdg \theta^a).
\ee
Using Eq. (\ref{cotton-york}), one can derive the tensorial expression
\be
C^{ab}
=
\epsilon^{acd}\nabla_c\left(R^b_{\fant{b}d}-\frac{1}{4}\delta^{b}_{d}R\right)
\ee
where $\epsilon^{acd}$ is the completely antisymmetric permuation symbol in three dimensions and  $\nabla_c$ stands for the covariant derivative in the expression.

\section{Field equations in the differential forms language}

In order to make  use of the null coframe formalism briefly developed in the previous section, it is essential to formulate the CGMG field equations in terms of differential forms.
For this purpose, the metric  field equations obtained by the coframe  variational derivative  can conveniently be expressed in terms of the 1-forms $E_a=E_{ab}\theta^b$. Explicitly, the fourth order field equations can be written as a 2-form equation  in the form \cite{baykal}
$
*E^a
\equiv
-\frac{1}{2}\delta L_{CGMG}/\delta \theta_a=0
$
with the explicit expression
\be\label{CGMG-eqns}
*E^a
=
\sigma *G^a+m^2\lambda *\theta^a+\frac{1}{\mu}C^a-\frac{1}{m^2}\left(D*C^a+*T^a[\Omega]\right)
\ee
where the quadratic-curvature part $*T_a[\Omega]=T_{ab}[\Omega]*\theta^b$ reads
\begin{align}
*T^a[\Omega]
&\equiv
\Omega_{bc}\wdg i^a*\left[\theta^b\wdg \left(R^c-\frac{3}{8}R\theta^c\right)\right]
\nonumber\\
&\qquad-
\frac{1}{2}\left(R_{bc}R^{bc}-\frac{3}{8}R^2\right)*\theta^a.
\end{align}

The Cotton 2-forms are traceless, $i_aC^a=0$, and the fourth order terms above, $D*C^a$, do not contribute to the trace of the metric field equations
as a consequence of the relation  $\theta_a\wdg D*C^a=D*i_aC^a=0$.
The second order trace is an important feature of the NMG Lagrangian that eliminates  the propagating spin-0 modes of the NMG model linearized around Minkowski background (for $\lambda=0$). Similarly, it is also well known that because $C^a$ satisfies $\theta^a\wdg C_a=0$, there is no third order contribution to the trace arising from the TMG part either. Moreover, the trace of the $*T^a[\Omega]$ gives back the quadratic curvature part of the CGMG Lagrangian,
\be
\theta_a\wdg *T^a[\Omega]
=
-\frac{1}{2}\left(R_{ab}R^{ab}-\frac{3}{8}R^2\right)*1,
\ee
unlike the case in four dimensions where the corresponding trace expression for $*T^a[\Omega]$ vanishes identically.

The field equations given by Eq. (\ref{CGMG-eqns}) for the CGMG model are valid  relative to  both an orthonormal coframe \cite{baykal,baykal2} as well as to the seminull coframe to be defined in the preliminary section above.

As an  illustration of  the expediency of the differential forms language, let us briefly discuss the linearization of the field equations (\ref{CGMG-eqns}) as a 2-form equation around the Minkowski background \cite{baykal-dereli-epjp}. By using the field equations (\ref{CGMG-eqns}) with $\lambda=0$ and  ignoring the Cotton part temporarily for the sake of simplicity of the argument, the linearization of the NMG equations can readily be written as a 2-form equation in the Minkowski spacetime as
\be\label{lin-2form-NMG}
d\star C^a_L-\sigma m^2 \star G^a_L=0,
\ee
with $R_L=0$ identically. Note that the linearized equations (\ref{lin-2form-NMG}) are readily obtained by the formal changes in (\ref{CGMG-eqns}). The label $L$ refers to the tensor-valued forms and tensor components linearized around the Minkowski spacetime. For example,
$G^a_L\equiv (G^{a}_{\fant{a}b})_Ldx^b$, $C^a_L\equiv\frac{1}{2}(C^{a}_{\fant{a}bc})_Ldx^b\wdg dx^c=d(Y^a)_L$ for the Einstein and the Cotton forms
respectively. Likewise, $\star$ is the Hodge dual in the Minkowski spacetime. With the help of the linearized Bianchi identity,
$(D*G^a)_L=d\star G^a_L=0$, the linearized 2-form equation, namely, Eq. (\ref{lin-2form-NMG}) can be rewritten in the familiar component form as
\be
(\Box+\sigma m^2)R^{ab}_L=0
\ee
which is an equation for a massive spin-2 field propagating in three dimensions \cite{BHT2} (Here $\Box\equiv \eta^{\mu\nu}\partial_\mu\partial_\nu$ in the flat
background). With $R_L=0$, the linearized Bianchi identity emulates the subsidiary condition $\partial_a R^{a}_{L b=0}$  on the massive spin-2 field.
By construction, for the pp-wave metric defined on Minkowski background, Eq. (\ref{lin-2form-NMG}) is equal to the
exact equations: $(*E^a)_L= E^a_{Lb}\star dx^b$ up to a term having second order linear in the derivatives of the metric coefficients, provided that
the background metric is chosen appropriately.  

Similarly, for $\sigma=-1$ and $\lambda=0$, the linearized CGMG equations can be written as a 2-form equation as
\be\label{linearized-GMG}
(\star d-m_+)(\star d+m_-)R^a_L=0
\ee 
with the masses $m_{\mp}$ defined in terms of the mass parameters of CGMG as 
\be\label{mass-redef}
\frac{1}{\mu}=\frac{1}{m_+}-\frac{1}{m_-}\quad \mbox{and}\quad m^2=m_+m_-.
\ee

The approximate expression (\ref{linearized-GMG}) is a 2-form equation concisely exhibiting  the claim that ``TMG is square root of NMG" at the linearized level.
(\ref{linearized-GMG}) is uplifted to  the level of the exact field equations  (\ref{CGMG-eqns}) in \cite{AA1} as a solution-generating technique for the  NMG model in connection with the solutions of TMG. In the case of CGMG, such a relation can be obtained by the mere rearrangement of the terms in 1-form $E^a$, which is defined 
in (\ref{CGMG-eqns}), as
\be
\left[*D *D-\frac{m^2}{\mu}*D+\sigma m^2\right]L^a=\tau^a
\ee
with the 1-form $\tau^a$ c defined as
\begin{eqnarray}
&&\tau^a
\equiv
*\left[\theta^{a}\wdg \theta^b\wdg \left(R^c-\frac{3}{8}R\theta^c\right)\right]*\Omega_{bc}
\nonumber\\
&&\phantom{aa}
+\frac{1}{2}\left[
\left(R_{bc}R^{bc}-\frac{3}{8}R^2\right)
+
2m^4\lambda-\frac{\sigma}{2} m^2R
\right]\theta^a.
\end{eqnarray}
In the same spirit as in the original formulation \cite{AA1}, which can be obtained as $\mu\rightarrow\infty$, it seems to be appropriate to write the CGMG equations in a tensorial ``Klein-Gordon" -type equation in terms of the Schouten 1-form in the formulation above. 

\section{The metric ansatz}

In terms of the local coordinates $\{x^\alpha\}=\{u, v, y\}$, where $u,v$ are real and null, and $y$ is a spatial coordinate,
the metric ansatz to be considered explicitly reads
\be\label{ansatz}
g=\frac{-2dudv-2H(u,y)du^2+dy^2}{\left[1+\frac{\Lambda}{4}(-2uv+y^2)\right]^2},
\ee
in a form conformal to the pp-wave metric. The parameter $\Lambda$, which determines the geometry of the background metric, admits arbitrary values and for $\Lambda=0$ the  pp-wave metric is recovered. For the construction of the new solutions, it is essential to have the profile function of the particular form $H(u,y)=\delta(u) h(y)$ involving a Dirac  delta  distribution function.

A seminull coframe basis that one can adopt is of the form
\be\label{coframe-basis}
k=P^{-1}{du},\quad l=P^{-1}(dv+Hdu),\quad m=P^{-1}dy,
\ee
where $P=1+\frac{\Lambda}{4}(-2uv+y^2)$
and thereby the ansatz (\ref{ansatz}) takes the desired form (\ref{metric-general-seminull-form}).
The frame fields associated to the above coframe can be constructed by making use of the metric duals of the coframe 1-forms defined in (\ref{coframe-basis}) above which explicitly read
\be
\tilde{l}=-P(\partial_u-H\partial_v),\quad \tilde{k}=-P\partial_v,\quad \tilde{m}=P\partial_y.
\ee
The null vector field associated to the basis 1-form $k$ is  defined by $\tilde{k}\equiv k^\alpha\partial_\alpha$
with $k^\alpha\equiv g^{\alpha\beta}k_\beta$ and $\tilde{k}$ satisfies the geodesic equation of the form
$
\nabla_{\tilde{k}} k=\frac{\Lambda}{2}uk
$
so that it is not affinely parameterized. 

In three dimensions, the optical scalars for a null geodesic vector field are defined in such a way that it differs from its four-dimensional counterpart. For the null geodesics, the shear and twist cannot be defined because the vector space $k^\perp/k$, that is the vector space defined as the orthogonal complement of the vector field $k$ quotiened by $k$ itself, is one-dimensional and is spanned by $m$ \cite{oneill}. Thus, the only optical scalar that can be defined on the vector space $k^\perp/k$ for a null geodesic vector field in the three-dimensional case is the divergence \cite{nurowski-chabert}. The particular metric ansatz (\ref{ansatz}) belongs to the Kundt family of metrics \cite{chow-pope-sezgin1,chow-pope-sezgin2} defined by a divergence-free, null geodesic vector field  in the general form for $\Lambda \neq0$.

In four dimensions, Siklos \cite{siklos} has shown that the only gravitational waves conformal to the pp-waves are the AdS-waves  corresponding to
the negative values of $\Lambda$. On the other hand, the impulsive wave case was later shown by Podolsk\`y  and Griffiths \cite{podolsky-griffiths, podolsky-griffiths-proof,siklos} to be an exception  to the Siklos' result. 
In three dimensions, the metric of the form  (\ref{ansatz}) with $\Lambda>0$ furnishes a new example  for the general result obtained by Ahmedov and Aliev \cite{AA3} asserting that all Petrov-Segre type N solutions to the NMG belong to the  Kundt family of metrics.

\section{The curvature forms}

Once the exterior derivatives of the basis 1-forms are calculated, the structure equations (\ref{se1-1}-\ref{se1-3}) can be solved for
the connection 1-forms to obtain
\begin{eqnarray}
&&\omega^{0}_{\fant{a}0}
=
\frac{\Lambda }{2}(vk-ul),
\qquad \omega^{0}_{\fant{a}2}
=
\frac{\Lambda }{2}(-yk+um),
\\
&&\omega^{1}_{\fant{a}2}
=
\frac{\Lambda}{2}(-yl+vm)+PH'k,
\end{eqnarray}
where a prime denotes a derivative with respect to the coordinate $y$.

Consequently, by inserting the connection expressions into the second structure equations (\ref{se2-1}-\ref{se2-3}), one finds
\begin{eqnarray}
&&\Omega^{0}_{\fant{a}0}
=
-\Lambda k\wdg l,
\qquad \Omega^{0}_{\fant{a}2}
=
\Lambda k\wdg m,
\\
&&\Omega^{1}_{\fant{a}2}
=
-\Phi(u,y)k\wdg m
+
\Lambda l\wdg m,\label{curvature-expressions}
\end{eqnarray}
where the function $\Phi(u,y)$ introduced above has the form
\be\label{phi-def}
\Phi(u,y)
=
P(PH''-P'H')+ \frac{\Lambda}{2}PH.
\ee

Relative to a coordinate basis defined by the components with, for example, $k=k_\alpha dx^\alpha$, the  traceless Ricci tensor  $S_{ab}=R_{ab}-\frac{1}{3}\eta_{ab}R$
has the canonical form $S_{\alpha\beta}=\Phi k_\alpha k_\beta$.  Although for $H=0$ and $\Lambda\neq0$ the background metric is of type O, for $\Lambda=0$ the metric  ansatz (\ref{ansatz}) is  Petrov-Segre type N  \cite{chow-pope-sezgin1} which can be inferred simply by examining the following expressions for the Ricci 1-forms:
\be
R^0=2\Lambda k,\quad R^1=-\Phi k+2\Lambda l,\quad
R^2=2\Lambda m.
\ee
It follows from these expressions that the scalar curvature $R=6\Lambda$ and that $\Phi$ can also be expressed in the form $\Phi=l^\alpha l^\beta R_{\alpha\beta}$.

The Cotton 2-forms can be calculated as
\be
C^0=0=C^2,\qquad 
C^1=\left(P\Phi'-P'\Phi\right)*k,
\ee
by using $C^a=DY^a$ and, also taking the important relation $u\Phi(u,y)\propto u\delta(u)\equiv0$ into account.

 These results allow one to write the TMG equations, $\mu^{-1}C^a+*G^a=0$, explicitly in the form
\be\label{TMG-eqns}
\left(P\Phi'-P'\Phi\right)+\mu \Phi=0.
\ee

The expressions for the Cotton 2-forms are then used to calculate that the fourth-order terms in the CGMG equations and the only
nonvanishing component  turns out to be 
\be\label{fourth-order-part}
D*C^1
=
\left[
P\left(P\Phi'-P'\Phi\right)'
-
P'\left(P\Phi'-P'\Phi\right)
\right]*k
\ee
by making use of the formula
\be
D*C^1=d*C^1-\omega^{0}_{\fant{a}0}\wdg *C^1+\omega^{1}_{\fant{a}2}\wdg *C^2.
\ee

Finally, one can show that the $*T^a[\Omega]$ has the explicit form
\begin{align}
*T^0[\Omega]
&=
-\frac{\Lambda^2}{4}*k,
\\
*T^1[\Omega]
&=
\frac{\Lambda}{4}\Phi*k
-
\frac{\Lambda^2}{4}*l,
\\
*T^2[\Omega]
&=
-\frac{\Lambda^2}{4}*m,
\end{align}
and, consequently, the  diagonal components $E^{0}_{\fant{a}0}=E^{1}_{\fant{a}1}=E^{2}_{\fant{a}2}$ yield an algebraic equation for $\Lambda$, 
\be\label{algebraic-NMG-eqn}
\frac{1}{4}\Lambda^2-\sigma m^2\Lambda+m^4\lambda=0,
\ee
relating the  mass and cosmological constant parameters of the CGMG model to the background curvature parameter.

As is the case for general quadratic curvature models in arbitrary dimensions, for $H=0$, the conformal 2-forms $C^a$ vanish identically and the decoupled equation 
(\ref{algebraic-NMG-eqn}) determines the maximally symmetric vacua for the NMG model which implies that the NMG model admits dS, AdS and flat vacuum solutions:
\be\label{lambda-values}
\Lambda_{\mp}
=
2m^2(\sigma\mp\sqrt{1-\lambda}).
\ee
Note, however, that $*T^1[\Omega]$ also contributes a term which is linear in $\Phi$ to the $*E^{1}_{\fant{1}0}$ component as well.

\section{The equation for the profile function}

By combining the  results of the calculations in the previous section, the fourth order linear differential equation for the profile function
obtained  from $E^1_{\fant{1}0}=0$ can be written in a factorized form as

\be\label{factorized-profile-eqn}
\left(\mathcal{D}-m_+\right)\left(\mathcal{D}+m_-\right)\left[p^4\left(\frac{1}{p}\frac{d}{dy}\right)^2+\frac{\Lambda p}{2}\right]h=0.
\ee
In the profile equation, the mass parameters $m_{\mp}$ are defined in Eq. (\ref{mass-redef}). $\mathcal{D}$ stands for the differential operator
\be\label{cal-derivative}
\mathcal{D}
\equiv
p^2\frac{d}{dy}\frac{1}{p}
\ee
with  
\be\label{small-p-def}
p\equiv
1+\frac{\Lambda}{4}y^2,
\ee
and $\Lambda$ parameter assumes the values determined by Eq. (\ref{lambda-values}). 

Moreover, the mass parameters $m_{\mp}$ in Eq. (\ref{factorized-profile-eqn}), which are defined  previously for flat background in (\ref{mass-redef}), are now defined by the relations of the form
\be\label{mass-redef2}
m_+-m_-=\frac{m^2}{\mu},\qquad m_+m_-=\sigma m^2+\frac{\Lambda}{4}
\ee
for a general curved background for the CGMG model.

The equation (\ref{factorized-profile-eqn}) for the profile function is to be compared with the general ``Klein-Gordon" form of the field equations given in (\ref{CGMG-eqns}). Furthermore, note that (\ref{factorized-profile-eqn}) has also formal resemblance to the linearized form of CGMG field equations (\ref{linearized-GMG}) with the exterior derivative replaced by the differential operator (\ref{cal-derivative}). For the flat background, the differential operator $\mathcal{D}$ reduces to a derivative with respect to the coordinate $y$.

The differential operators $\mathcal{D}-m_+$ and $\mathcal{D}+m_-$ commute and the general solution of the second order equation
\be
\left(\mathcal{D}-m_+\right)\left(\mathcal{D}+m_-\right)\Phi=0
\ee 
can be written in the form $\Phi=\Phi_{+}+\Phi_-$  with the functions $\Phi_{\mp}$ being solutions to the equations
\be
\left(\mathcal{D}-m_+\right)\Phi_+=0,
\qquad
\left(\mathcal{D}+m_-\right)\Phi_-=0.
\ee
One readily finds that
\be
\Phi_+
=
C_+p^{(1+m_+)},
\qquad
\Phi_-
=
C_-p^{(1-m_-)},
\ee
where $C_{\mp}$ are integration constants. Consequently, the fourth order equation for the profile function (\ref{factorized-profile-eqn}) reduces to a second order inhomogeneous equation of the form
\be\label{reduced-second-order-de}
ph''
+
\frac{\Lambda }{2}
\left(
-yh'
+
h
\right)
=
C_+p^{m_+}
+
C_-p^{-m_-}
\ee
with the function $p(y)$ defined as in (\ref{small-p-def}).

Note that the term on the left-hand side in (\ref{reduced-second-order-de}) is the Einstein tensor  $G^{1}_{\fant{a}0}$ 
in three dimensions (cf., Eqs. (\ref{3d-einstein-forms}) and (\ref{curvature-expressions})) up to an overall factor $p$. Therefore, the factorization introduced in (\ref{factorized-profile-eqn}) eventually leads to an equation for the profile function of an impulsive gravitational wave in Einstein gravity in three spacetime dimensions  with an effective source term arising from the higher order terms in the CGMG model. In other words, the gravitational wave ansatz (\ref{ansatz}) reduces the general the CGMG field equations in (\ref{CGMG-eqns}) to those of three dimensional Einstein field equations with an effective source term depending on the mass parameters $m_{\mp}$ in the CGMG model  defined in (\ref{mass-redef2}) as well as the background curvature parameter $\Lambda$ determined by Eq. (\ref{lambda-values}).  Consequently, the particular solution $h_p(y)$ of the reduced equation (\ref{reduced-second-order-de}) is of interest for the impulsive wave solutions of CGMG theory and $h_p(y)$ can be written as a superposition of two particular solutions $\Phi_{\mp}(y)$.

The solutions to the homogeneous equation for the differential equation (\ref{reduced-second-order-de}) around the origin can be found as follows.
Because $y=0$ is an ordinary point of the differential equation, the analytical solution assumes  the form of an infinite series 
$h=\sum_{k=0}^{\infty}c_ky^k$. By inserting the series expression into the homogeneous equation, one finds that the coefficients 
satisfy the following recurrence relation
\be
c_{k+2}
=
-\frac{\Lambda}{4}\frac{(k-1)(k-2)}{(k+1)(k+2)}c_k,
\ee
which is valid for $k\geq0$. Consequently, both the odd and even power series in the general solution truncate and the independent homogeneous solutions, 
denoted by $h_1(y)$ and $h_2(y)$  take the form
\be\label{hom-sol}
h_1(y)=y,\qquad h_2(y)=1-\frac{\Lambda}{4}y^2,
\ee
respectively.

The particular solution can be obtained by using the homogeneous solutions to construct the corresponding Green's function (see, for example,  \cite{dennery}). Explicitly, the particular solution can be expressed in terms of the Green's function as an integral of the form
\be\label{particular-sol}
h_p(y)
=
\int^{y}G(y,\xi)p^{-3}(\xi)\left[\Phi_+(\xi)+\Phi_-(\xi)\right]d\xi,
\ee
where $\xi$ is a continuous parameter on the $y$ axis.

The Green's function can be decomposed  into  singular and homogeneous parts as 
\be
G(y,\xi)
=
G_s(y,\xi)+G_h(y,\xi),
\ee
 where the homogeneous part is a linear superposition of the homogeneous solutions that satisfy some given boundary conditions on the general solution. 

In terms of the set of homogeneous solutions, the singular part of the Green's function  can be written  as
\be
G_{s}(y,\xi)
=
\frac{h_1(\xi)h_2(y)-h_1(y)h_2(\xi)}{p^{-1}(\xi)W[h_1(\xi),h_2(\xi)]}\theta(y-\xi),
\ee
where $\theta$ stands for the unit step function, and $W[h_1, h_2]$ is the Wronskian.
In accordance with these definitions, $G_{s}(y,\xi)$ satisfies the equation 
\be\label{hermitian-do-def}
\left(
\frac{d}{dy}\frac{1}{p}\frac{d}{dy}
+
\frac{\Lambda}{2p^2}
\right)
G_{s}(y,\xi)=\delta(y-\xi).
\ee
The differential operator on the left-hand side in (\ref{hermitian-do-def}) is a Hermitian operator which  can be obtained from the differential equation in 
(\ref{reduced-second-order-de}) by multiplying it with $p^{-2}$. 

By using the explicit expressions for the homogeneous solutions, one can show that $G_{s}(y,\xi)$ 
takes the form  
\be
G_{s}(y,\xi)
=
\left(\frac{1+\frac{\Lambda}{4}\xi^2}{1-\frac{3\Lambda}{4}\xi^2}\right)\left(1-\frac{\Lambda}{4}y\xi\right)(y-\xi)\theta(y-\xi).
\ee

The linear superposition of the solutions in (\ref{hom-sol}) and (\ref{particular-sol}) constitutes the most general
solution for the profile function $h(y)$ in the neighbourhood of the point $y=0$ in closed form for the impulsive gravitational waves in CGMG theory.

Finally, we note that in finding an explicit expression for a solution to Eq. (\ref{reduced-second-order-de}), the cases $\Lambda>0$ and $\Lambda<0$ are to be handled separately. For example, in the case  $\Lambda<0$, the points $y=\pm1$ are regular singular points of the corresponding equation, and therefore the validity for the general solution discussed above is limited to the range $|y|<1$.

\section{Concluding comments}

It is worth  emphasizing that the construction  of the new solutions depends crucially on the assumption that the profile function has a Dirac-delta function distribution factor and it does not work otherwise. The general solution to the reduced equation (\ref{reduced-second-order-de}) does not belong to  the universal class of solutions presented in \cite{AA3} obtained by the analytical continuation of the parameters of the TMG/NMG model.  

Ahmedov and Aliev \cite{AA1} showed that all Petrov-Segre type D and N solutions of TMG can be mapped to the solutions of NMG by rewriting the 
NMG and TMG field equations in terms of  a covariant differential operator, denoted by $\slashed{D}$ in the original notation. 
With the application of the covariant operator $\slashed{D}$ to TMG equations, one ends up with NMG field equations  under certain assumption  on the traceless Ricci tensor. In this exact sense, the TMG field equations can be considered as square root of NMG field equations.
In this regard, the above reduction of the CGMG field equations to a three-dimensional Einstein field equations (\ref{reduced-second-order-de}) 
and a constraint equation (\ref{lambda-values}) for the particular case of the impulsive metric ansatz is an interesting result for the CGMG theory that  deserves further scrutiny from a broader point of view, possibly by considering a whole family of metrics belonging to a particular Petrov-Segre type.

The impulsive gravitational waves with null particle sources in four spacetime dimensions have previously been studied by Podolsk\'y and Griffiths
\cite{podolsky-griffiths-proof} extending the well known work of Aichelburg  and Sexl \cite{aichelburg-sexl} to dS and AdS backgrounds. 
In a more recent work \cite{edelstein-tekin}, introducing impulsive gravitational  waves generated by null  particles
in three spacetime dimensions \cite{deser-mccharty-steif}, it has been shown that the local causality and unitarity (the absence of ghosts and tachyons at the linearized level) are not in conflict in flat and AdS backgrounds for massive gravity models. 

Penrose \cite{penrose-cut-paste} constructed impulsive waves  by a geometrical method which is known as  Penrose's Cut and Paste method \cite{exact-sol-GP}. 
The ``cut and paste" method is later extended to the  impulsive waves to include an arbitrary cosmological constant by Podolsk\'y and Griffiths \cite{podolsky-griffiths-proof,podolsky-griffiths}. In a similar vein, the present work constructs impulsive wave solutions in higher curvature massive models in three  dimensions.

Although the pp-wave ansatz  linearizes the CGMG equations in a maximally symmetric curved background for $\Lambda\neq0$, the resulting field equations are not as simple as they are for the flat Minkowski background spacetime. By construction, the interpretation of the massive spin-2 field by the NMG model crucially depends on the flat background although the vacuum field equations admit  dS/AdS solutions as well.  On the other hand, it is well known that in a constant curvature background  massive spin-2 has fewer degrees of freedom then that of the usual massive spin-2 case 
\cite{carlip-deser-waldron-wise,deser-nepomechie,deser-steif,btekin,hinterbichler}. Moreover, the subsidiary  conditions, which are essential to have a consistent, free, massive spin-2 field interpretation, would differ from those that follow from linearization around  dS or AdS backgrounds.

\emph{We dedicate our work to the memory of  Professor John Freely.}




\end{document}